\documentclass[aps,twocolumn,,superscriptaddress,pacs]{revtex4-1}
\usepackage{graphicx}
\usepackage{bm, amsmath, amssymb}
\usepackage{color}
\usepackage{times}
\newcommand{\beq}{\begin{equation}}
\newcommand{\eeq}{\end{equation}}
  \newcommand{\beql}[1]{\begin{equation}\label{eq:#1}}
  \newcommand{\beqa}{\begin{eqnarray}}
  \newcommand{\eeqa}{\end{eqnarray}}
  \newcommand{\M}{{\bf M}}
 \newcommand{\bM}{\mathbf{M}}
  \newcommand{\bP}{\mathbf{P}}
  \newcommand{\bS}{\mathbf{S}}
  \newcommand{\da}{\dagger}
  \newcommand{\de}{\delta}
  \newcommand{\ep}{\varepsilon}
  \newcommand{\et}{\eta}
  \newcommand{\la}{\lambda}
  \newcommand{\mb}{\mbox}
  \newcommand{\nn}{\nonumber}
  \newcommand{\si}{\sigma} 
  \newcommand{\ta}{\tau}
                                            
 \newcommand{\Eq}[1]{Eq.~(\ref{eq:#1})}                                  
  \newcommand{\Tr}{\mbox{\rm Tr}}
  \newcommand{\eq}[1]{(\ref{eq:#1})}
\newcommand{\bra}[1]{\langle#1|}
\newcommand{\ket}[1]{|#1\rangle}

\newcommand{\ketbra}[1]{\ket{#1}\bra{#1}}
\renewcommand{\t}{t_0}
\newcommand{\av}[1]{\langle #1 \rangle}

\newcommand{\bmat}{\left(\begin{array}{cc}}
\newcommand{\emat}{\end{array}\right)}
\newcommand{\bvec}{\left(\begin{array}{r}}
\newcommand{\evec}{\end{array}\right)}

\def\1{\mathchoice{\rm 1\mskip-4.2mu l}{\rm 1\mskip-4.2mu l}{\rm
        1\mskip-4.6mu l}{\rm 1\mskip-5.2mu l}}
\newcommand{\deq}[1]{\begin{align}#1\end{align}}
\newcommand{\deqs}[1]{\begin{align*}#1\end{align*}}
\newcommand{\Det}{\Delta t}
\newcommand{\hep}{\hat{\varepsilon}}
\newcommand{\het}{\hat{\eta}}

\begin{document}
\title{Violation of Heisenberg's error--disturbance relation by Stern--Gerlach measurements}
\author{Yuki Inoue}
\email[]{inoue.y.at@gmail.com}
\affiliation{Graduate School of Informatics, Nagoya University,
Chikusa-ku, Nagoya, 464-8601, Japan}
\author{Masanao Ozawa}
\email[]{ ozawa@is.nagoya-u.ac.jp}
\affiliation{Graduate School of Informatics, Nagoya University,
Chikusa-ku, Nagoya, 464-8601, Japan}
\affiliation{College of Engineering, Chubu University,
1200 Matsumoto-cho, Kasugai-shi, Aichi, 487-8501, Japan}

\begin{abstract}
Although Heisenberg's uncertainty principle is represented by a rigorously proven relation about intrinsic indeterminacy in quantum states, Heisenberg's error--disturbance relation (EDR) has been commonly believed as another aspect of the principle. However, recent developments of quantum measurement theory made Heisenberg's EDR testable to observe its violations. 
Here, we study the EDR for Stern--Gerlach measurements.
In a previous report, it has been pointed out that their EDR is close to the theoretical optimal.  The present note reports that even the original Stern--Gerlach experiment in 1922, the available experimental data show, violates Heisenberg's EDR. 
The results suggest that Heisenberg's EDR is more ubiquitously violated than it has long been supposed. 
\end{abstract}


\maketitle
\section{Introduction}
Heisenberg's uncertainty principle is usually represented by a rigorously proven relation
\beql{Rob29}
\si(A)\si(B)\ge\frac{1}{2}|\av{[A,B]}|
\eeq
for the standard deviations $\si(A),\si(B)$ of arbitrary observables $A,B$, respectively,
in any state \cite{Hei27,Ken27,Rob29}.  This expresses intrinsic indeterminacy in quantum states. 
However, Heisenberg's error--disturbance relation (EDR)
\beql{Hei27}
\ep(A)\et(B)\ge\frac{1}{2}|\av{[A,B]}|
\eeq
for the mean error $\ep(A)$ of an $A$-measurement in any state 
and the mean disturbance $\et(B)$ thereby caused on another observable $B$,
originally introduced his $\gamma$-ray microscope thought experiment \cite{Hei27}, 
has been commonly believed and taught as another aspect of the principle.  
Although no general proofs have been known,
there have been continuing efforts to prove Heisenberg's EDR \eq{Hei27}, which result  in 
proving \Eq{Hei27} for jointly unbiased measurements  \cite{AK65,YH86, AG88,Ish91,91QU}
(in a wider context of approximate simultaneous measurements)
and measurements with independent interventions \cite{03UVR,04URN}.
However, recent developments of quantum measurement theory derived a universally 
valid EDR
\beql{Oza03}
\ep(A)\et(B)+\ep(A)\si(B)+\si(A)\et(B)\ge\frac{1}{2}|\av{[A,B]}|,
\eeq
where $\si(A)$ and $\si(B)$ are the standard deviations of $A$ and $B$ just before the
measurement \cite{03UVR,04URN},
and made Heisenberg's EDR testable to observe its experimental violation \cite{12EDU,13VHE}.
Subsequently, stronger EDRs have appeared \cite{Bra13,14EDR} and further
experimental violations of Heisenberg's EDR have been reported, 
though witnessed only in ideally controlled precision measurements of
photons  \cite{LW10,RDMHSS12,13EVR,WHPWP13,14A1,RBBFBW14}
and neutrons  \cite{16A3}.

Here, we study the EDR for a more common measurement setup, known as  
Stern--Gerlach measurements \cite{SLB87,CG05,PBCG05,HPAM07,Dev15}.
In a previous report \cite{IO20}, it has been pointed out that their 
 error--disturbance region is close to the theoretical optimal and that the Heisenberg's EDR 
can be violated in a broad range of experimental parameters. 
The present note reports that in fact, the available experimental data show,
the original Stern--Gerlach experiment performed in 1922 \cite{GS22a,GS22b,GS22c} 
violates Heisenberg's EDR.
The results suggest that Heisenberg's EDR is more ubiquitously violated than it has been 
supposed for a long time. 

\section{Spin measurements}
We consider a measurement of a spin-1/2 particle, $\bS$,
or an equivalent q-bit system described by Pauli matrices.
We investigate the error and disturbance of the measurements
of the $z$-component, $A=\si_z$, and the disturbance of the $x$-component, 
$B=\si_x$, of the (dimensionless) spin, where
$A$ and $B$ generally denote observables to be measured and
to be disturbed, respectively.
We suppose that the measurement is carried out by the interaction between
the system $\bS$  prepared in an arbitrary state $\rho$
and the probe $\bP$ prepared in a fixed vector state $\ket{\xi}$
from time 0 to time $\t$ and ends up with the subsequent reading of the meter 
observable $M$ of the probe $\bP$.
We assume the meter $M$ has the same spectral with the measured observable $\si_z$.
The measuring process, $\bM$, determines the time evolution operator, $U$, 
of the composite system of $\bS+\bP$.
For any observables $X$ in $\bS$ and $Y$ in $\bP$, the Heisenberg operators 
at the corresponding times are given by $X(0)=X\otimes \1$, $Y(0)=\1\otimes Y$,
$X(\t)=U^{\da}(X\otimes \1)U^{\da}$, and $Y(\t)=U^{\da}(\1\otimes Y)U^{\da}$.

The quantum root-mean-square (q-rms) error,
$\ep(\si_z)=\ep(\si_z,\bM,\rho)$,
is defined by
\begin{align}
\ep(\si_z)&=\Tr[(M(\t)-\si_z(0))^2\rho\otimes\ketbra{\xi}]^{1/2}.
\end{align}
The q-rms error $\ep(\si_z)$ has the following properties \cite{19A1}.

(i) (Operational definability) $\ep(\si_z)$ is definable by the operational description of 
the measuring process $\bM$.

(ii) (Correspondence principle) If $\si_z(0)$ and $M(\t)$ commute in the state
$\rho\otimes\ketbra{\xi}$, the q-rms error $\ep(\si_z)$ coincides with the classical rms error determined by
the joint probability distribution $\mu$ of $\si_z(0)$ and $M(\t)$ in $\rho\otimes\ketbra{\xi}$.

(iii) (Soundness) If $\bM$ accurately measures $\si_z$ in $\rho$ then $\varepsilon(\si_z)$ vanishes.

(iv) (Completeness) If $\varepsilon(\si_z)$ vanishes then $\bM$ accurately measures $\si_z$ in $\rho$.
 
The quantum root-mean-square (q-rms) disturbance,
$\et(\si_x)=\ep(\si_x,\bM,\rho)$, is defined by
\begin{align}
\et(\si_x)&=\Tr[(\si_x(\t)-\si_x(0))^2\rho\otimes\ketbra{\xi}]^{1/2}.
\end{align}
The q-rms disturbance  $\et(\si_x)$ has properties analogous to the q-rms error.

For the above properties of $\ep(\si_z)$ and $\et(\si_x)$, we refer the reader to 
Ref.~\cite{19A1} and Appendix A of Ref.~\cite{IO20}.

According to Braciard \cite{Bra13} and Ref.~\cite{14EDR},  we obtain the EDR
\begin{equation}
\hat \varepsilon (\si_z)^2  + \hat \eta (\si_x)^2+ 2  \hat \varepsilon(\si_z) \hat \eta(\si_x) \sqrt{1-D_{\si_z\si_x}^2}
\geq D_{\si_z\si_x}^2,
\label{eq:BOEDRS}
\end{equation}
where
$D_{AB}=\frac{1}{2}\mathrm{Tr}(\left| \sqrt{\rho }[A, B] \sqrt{\rho }\right|)$, 
\deqs{
\hep(A)=\ep(A)\sqrt{1-\frac{\ep(A)^2}{4}}, \mb{ and }
\het(B)=\et(B)\sqrt{1-\frac{\et(B)^2}{4}}.
}
In the case where
\begin{equation}
\left\langle \sigma _z \right\rangle _{\rho } = \left\langle \sigma _x \right\rangle _{\rho } = 0, 
\label{eq:condition0}
\end{equation}
relation \eq{BOEDRS} is reduced to the tight relation
\begin{equation}
\left(
 \varepsilon  (\sigma _z) ^2 -2
\right)^2
+
\left(
 \eta (\sigma _x) ^2 -2
\right)^2
\leq 4.
\label{eq:BOEDRSZ}
\end{equation}
See  Appendix A in Ref.~\cite{IO20}.

Lund and Wiseman \cite{LW10} proposed a measurement model $\M(\theta)$
measuring $\si_z$ of the system $\bS$
with another q-bit system as the probe $\bP$ prepared in the state 
$\ket{\xi(\theta)}=\cos\theta\ket{0}+\sin\theta\ket{1}$ with the meter observable 
$M=\si_z$ of the probe $\bP$.
The measuring interaction is described by the controlled-NOT (CNOT) operation
$U_{{\rm CNOT}}=\ketbra{0}\otimes\1+\ketbra{1}\otimes \si_x$. 
For any state $\rho$
the error
 $\ep(\si_z)$ 
and the disturbance $\et(\si_x)$ of $\M(\theta)$ 
satisfy
$\ep(\si_z)=2|\sin \theta|$ and
$\et(\si_x)=\sqrt{2}|\cos\theta-\sin\theta|$.
Thus, they attain the bound
\begin{align}
(\ep(\si_z)^2-2)^2+(\et(\si_x)^2-2)^2=4
\end{align}
for the tight EDR \eq{BOEDRSZ}.
Experimental realizations of this model were reported by Rozema et al. \cite{RDMHSS12} 
and Refs.~\cite{13EVR,WHPWP13,14A1,RBBFBW14,16A3}.

In this study, we consider another type of measurement model measuring  $\si_z$,
known as Stern--Gerlach measurements, and investigate the admissible region of the error 
$\ep(\si_z)$ for $\si_z$ measurement and the disturbance $\ep(\si_x)$ on $\si_x$,
obtained from Gaussian orbital states.  

\section{Stern--Gerlach Measurements}
\begin{figure}[h]
        \centering
        \includegraphics[width=0.40\textwidth]{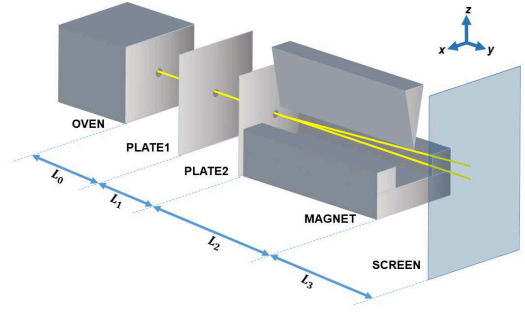}
    \caption{
    Illustration of the experimental setup for a Stern--Gerlach measurement \cite{IO20}. 
    The relations between the length and the time interval are $L_2 = v_y \Delta t , L_3 = v_y \tau $.
    }
    \label{sgexp}
\end{figure}

Let us consider the setting of a Stern--Gerlach measurement as depicted in Figure \ref{sgexp}. 
A particle with spin-$1/2$ goes through the inhomogeneous magnetic field and then evolves freely.
The inhomogeneous magnetic field is approximated to be
$
\mathbf{B} \simeq
\left(
 \begin{array}{ccc}
  0, & 0, & B_0 + B_1 z
 \end{array}
\right)
$.
The state of the spin degree of freedom $\bS$ is supposed to be an arbitrary mixed state satisfying 
$\left\langle \sigma _z  \right\rangle _{\rho } =\left\langle \sigma _x  \right\rangle _{\rho } = 0$,
e.g., $\rho=\ketbra{\si_y=\pm1}$.

The measuring process of this Stern--Gerlach measurement is given as follows.
The probe system $\bP$ is 
the $z$-component of the orbital degree of freedom of the particle. 
We assume that the initial state of the probe system $\bP$ is a general Gaussian state 
given by
$
\xi _\lambda(z)=A\exp \left( -\lambda z^2 \right)
$, 
where $\lambda \in \mathbb{C}$ and $\rm{Re}\, \lambda >0$.
The Hamiltonian of the composite system $\bS+\bP$ is given by
\begin{equation}
H(t)\!=\!\left\{\!\!
\begin{array}{lr}
\mu\sigma _z\otimes\left( B_0 + B_1 Z\right) 
 +\dfrac{1}{2m}\1\otimes P^2
 &
\hspace{-30pt} 
(0 \leq t \leq \Delta t ), \\
 \displaystyle\frac{1}{2m}\1\otimes P^2 
 & 
 \hspace{-30pt} 
(\Delta t \leq t \leq \Delta t + \tau).
\end{array}
 \right.
\end{equation}
The meter observable is
$
M= f(Z),
$
where
\begin{equation}
f(z)
=
\begin{cases}
-1 & (\mbox{if} \ z \geq 0), \\
+1 & (\mbox{if} \ z <0). 
\end{cases} \nonumber
\end{equation}

\section{Error and Disturbance in Stern--Gerlach Measurements}
Under the condition above, we obtain the following formulae for the error and disturbance 
in Stern--Gerlach measurements:
\deq{
&\ep(\si_z)^2=
2\,{\rm erfc}\left(
\frac{g_0}{\sqrt{2}\si(\Det+\ta)}
\right),
\label{eq:error}\\
&\eta(\sigma _x )^2=2\!-\!2
\exp \!
\left[
 -\frac{2 \mu ^2 B_1^2 \Delta t^2 }{\hbar ^2}
  \si\left(\frac{\Delta t}{2} \right)^2 
\right]
\cos \!
\frac{2\mu \Delta t B_0}{\hbar } ,\quad
\label{eq:disturbance}
}
where the  complementary error function, ${\rm erfc}(x)$, 
and the parameters $g_0$ and $\si(t)$ are given by 
\begin{align}
{\rm erfc}(x)&=\frac{2}{\sqrt{\pi}}
 \int_{x}^{\infty }
\exp(-w^2) dw,\\
g_0 &=
\frac{\mu B_1 \Delta t }{m}
\left(
\frac{\Delta t}{2}+ \tau 
\right),  \\
\si(t) &=
\left\langle
 \left(
  Z + \frac{t}{m}P
 \right)^2
\right\rangle_{\xi_\la}^{1/2}. 
\end{align}
See Eqs.~(62) and (69) in Ref.~\cite{IO20} for the detailed derivations.

The parameter $\si(\Det/2)$ represents the spread of the wave 
packet of the particle in the Stern--Gerlach magnet. 
The particle in the Stern--Gerlach magnet is exposed to the inhomogeneous magnetic field 
and its spin is precessed in an uncontrollable way. 
The parameter $\si(\Det/2)$ appears in the formula of the disturbance, because 
the disturbance of the spin along the $x$-axis is caused by 
this uncontrollable precession around $z$-axis. 
On the other hand, the error in the Stern--Gerlach setup comes from the non-zero dispersion $\si(\Det+\ta)$ of the particle position 
on the screen. By the uncertainty relation 
\deq{
\si\left(\frac{\Det}{2}\right)\si(\Det+\ta)\ge\frac{\hbar}{2m}\left(\dfrac{\Det}{2}+\ta\right),
}
the smaller the dispersion $\si(\Det+\ta)$ of the particle position 
on the screen, the greater the dispersion $\si(\Det/2)$  
of the the particle position in the Stern--Gerlach magnet.
This is why $\si(\Det+\ta)$ appears in the formula of the error, and this yields a tradeoff between 
$\ep(\si_z)$ and $\et(\si_x)$.

\section{Minimizing Error of Stern--Gerlach Measurements}
We minimize the error $\varepsilon (\sigma _z) $ 
with respect to the time interval $\tau $ of free evolution 
after leaving the magnetic field.
If the condition
\begin{equation}
m
\left\langle
 \left\{
  Z, P
 \right\}
\right\rangle_{\xi _{\lambda } }
+
\left\langle
 P^2
\right\rangle_{\xi _{\lambda } }
\Delta t
<0 \label{condition}
\end{equation}
holds, then the error is minimized 
at
\begin{align}
\tau
&= \tau_0  \notag  \\
&=
-\frac
{
 4m^2
 \left\langle
  Z^2
 \right\rangle _{\xi _{\lambda } }
 +
 3m
 \left\langle
  \left\{
   Z,  P
  \right\}
 \right\rangle_{\xi _{\lambda } }
 \Delta t
 +
 2 \left\langle P^2 \right\rangle_{\xi _{\lambda }} \Delta t ^2}
{ 2
 \left(
  m
  \left\langle
   \left\{
    Z, P
   \right\}
  \right\rangle_{\xi _{\lambda } }
  +
  \left\langle
   P^2
  \right\rangle_{\xi _{\lambda } }
  \Delta t
 \right)}.
\end{align}
Otherwise, the error is minimized as $\tau $ goes to infinity.
See Eq.~(65) in Ref.~\cite{IO20} for the detailed derivation.

\section{Range of the Error and Disturbance in Stern--Gerlach Measurements}
\begin{figure}[h]
\begin{center}
\includegraphics[width=0.45\textwidth]{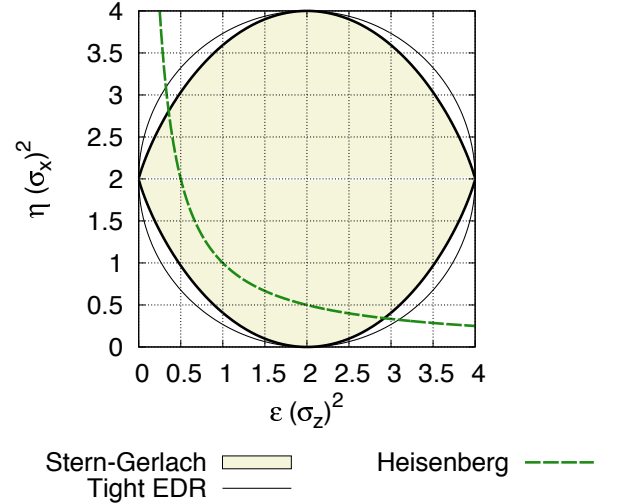}
\caption{
The range of the error and disturbance for Stern--Gerlach measurements \cite{IO20}. 
 Beige region: the region \eq{EDR-SG} that Stern--Gerlach experiment can achieve.
Black thine line: the boundary of the tight EDR \eq{BOEDRSZ}.
Green dashed  line: the boundary of Heisenberg's  EDR \eq{Hei27}.
}
\label{errordisturbancesg}
\end{center}
\end{figure}

We regard $\left( \varepsilon (\sigma _z ), \eta (\sigma _x ) \right)$ as a function of variables $\lambda , B_0 ,\tau >0$. 
As depicted in Figure \ref{errordisturbancesg}, 
the range of the function $\left( \varepsilon (\sigma _z ), \eta (\sigma _x ) \right)$ is obtained as 
\begin{equation}
\left|
 \frac{2-\eta (\sigma _x) ^2}{2}
\right|
\leq
\exp
\left\{
 -
 \left[
  \mathrm{erf}^{-1}
  \left(
   \frac{2-\varepsilon (\sigma _z)^2}{2}
  \right)
 \right] ^2
\right\}, 
\label{eq:EDR-SG}
\end{equation}
where 
$\mathrm{erf}^{-1}$
represents the inverse function of the error function 
$\mathrm{erf} (x) = \frac{2}{\sqrt{\pi }}\int _{0}^{x}\exp (-s^2)ds$.
See Eq.~(76) in Ref.~\cite{IO20} for the detailed derivation.

\section{Original Stern--Gerlach measurement} \label{originalSG}
Here, we estimate the error and disturbance of the original Stern--Gerlach experiment conducted by Stern and Gerlach
\cite{GS22a, GS22b,GS22c} 
by our theoretical model. 
We summarize the set up of their experiment (cf. Figure \ref{sgexp}).
A beam of silver atoms emerging from a small hole of a lid of an oven heated to 
$1500 \left[\mathrm{K}\right]$
 was collimated by two plates made of platinum. 
The atoms passed a pinhole with an area of $3\times 10^{-3} [\mathrm{mm}]$
(or $d_1 = 6.2 \times 10^{-2} \left[ \mathrm{mm} \right]$ in diameter) 
in the first plate P$_1$ and then passed the slit
$d_2 = 3.0 \mb{ to } 4.0 \times 10^{-2} \left[ \mathrm{mm} \right]$ in width in the second plate P$_2$. 
The slit was parallel to the $x$-axis. 
These plates were arranged perpendicular to the orbit of the atoms and the distance between them was $L_1 = 3.3 \left[ \mathrm{cm} \right]$. 
An $L_2 = 3.5 \left[ \mathrm{cm} \right]$ long knife edged magnetic pole was arranged parallel 
to the orbit of atoms just after the plate P$_2$. 
The $z$-component of the gradient of the magnetic field around the orbit of atoms was 
$B_1 = -1.35 \times 10^3 \left[ \mathrm{T} \cdot \mathrm{m}^{-1} \right]$. 
A glass plate was arranged immediately after the magnetic pole, in which the atoms are deposited. 
These conditions of the experiment is summarized in Table \ref{SGexptable}.

\begin{table}[h]
\caption{The data for the experiment conducted by Gerlach and Stern \cite{GS22a,GS22b,GS22c} 
in 1922.}
\bigskip

\begin{tabular}{c|c|c}
Experimental & & \\ Parameters & Values & Related Variables \\
\hline \hline
Temperature $T$ & & \\ of Oven        & $1500 \left[\mathrm{K}\right]$                                 & $\Delta t$, $\tau $ \\ \hline
Gradient $B_1$ of & & \\ Magnetic Field  & $ -1.35 \times 10^3 \left[\mathrm{T}/\mathrm{m}\right]$ & $B_1$ \\ \hline
$L_1$                                  & $3.3 \times 10^{-2}  \left[\mathrm{m}\right]$                & $\xi$ \\ \hline
$L_2$                                  & $3.5 \times 10^{-2} \left[\mathrm{m}\right]$                 & $\Delta t$ \\ \hline
$L_3$                                  & $0 \left[\mathrm{m}\right]$                                        & $\tau $\\ \hline
Diameter $d_1$ of & &\\ Hole of Plate1      & $6.2 \times 10^{-5}\left[\mathrm{m}\right]$                   & $\xi $\\ \hline
Width $d_2$ of Slit & &\\ of Plate2             & $4.0 \times 10^{-5}\left[\mathrm{m}\right]$                   & $\xi $
\end{tabular}
\label{SGexptable}
\end{table}

After the $8$ hours of the operation of the system and developing, they obtained a lip-shaped pattern. 
The maximum width of the opening of the lip shaped pattern was $1.1 \times 10^{-1} \left[ \mathrm{mm} \right]$.
The distance between the centers of the two arc-shaped pattern was $2.0 \times 10^{-1} \left[ \mathrm{mm} \right]$.
The velocity distribution of atoms in the oven is assumed to be the Maxwell distribution. 
Thus, the atoms emerging from the small hole of the lid of the oven are estimated to have the well-known distribution 
of flux \cite{Ste20b}:
\begin{equation}
f_{\mathrm{flux}} (v) = \mathrm{Const}. \times v^3 \exp \left( - \frac{mv^2}{2 k_B T} \right). \label{flux}
\end{equation}
The the root-mean-square $v_y$ of the $y$-component of the velocity of atoms is given by \cite{Ste20b}
\begin{equation}\label{eq:v_y}
v_y =
\sqrt{
\frac{4k_B T}{m}
} .
\end{equation}

Let us estimate the $z$-component $\ket{\xi_\la}$ of the orbital state of an atom in the beam just 
before entering the magnetic field. 
We assume the orbital state arriving at plate 1 to be 
$\xi_a(z)=(2a/\pi)^{1/4}\exp(-a z^2)$ with $a>0$.
We model the operations of the collimator and the slit 
as approximate momentum-position successive measurements 
by the canonical $D_p$-approximate momentum measurement and 
the canonical $D_z$-approximate position measurement 
introduced in \cite[Eq.~(75)]{93CA},
so that for the outcomes $(P,Z)=(0,0)$ the posteriori (output) state $\ket{\xi_\la}$ for the prior (input) 
state $\ket{\xi_a}$ is given by
\deq{
\ket{\xi_\la}\propto\exp\left(-\frac{Z^2}{4D_z^2}\right)\exp\left(-\frac{P^2}{4D_p^2}\right)\ket{\xi_a},
}
where $\propto$ stands for the equality up to a constant factor.
The parameters $D_p$ and $D_z$ will later be determined relative to the structure of the collimator
and the slit. 
Then, we have
\deq{
\xi_{\la}(z)\propto\exp\left\{
-\left[\left(\frac{1}{a}+\frac{\hbar^2}{D_p^2}\right)^{-1}+\frac{1}{4D^2_z}\right]
z^2\right\}.
}
We naturally assume $\si(P)_{\xi_a}\gg D_p$, so that we have
\deq{
\frac{1}{a}=4\si(Z)^2_{\xi_a}=\frac{\hbar^2}{\si(P)^2_{\xi_a}}\ll \frac{\hbar^2}{D_p^2}
}
and we have
\deq{
\xi_{\la}(z)\propto\exp\left[
-\left(\frac{D_p^2}{\hbar^2}+\frac{1}{4D^2_z}\right)
z^2\right]
}
up to arbitrary order.

\begin{figure}[h]
\begin{center}
\includegraphics[width=0.45\textwidth]{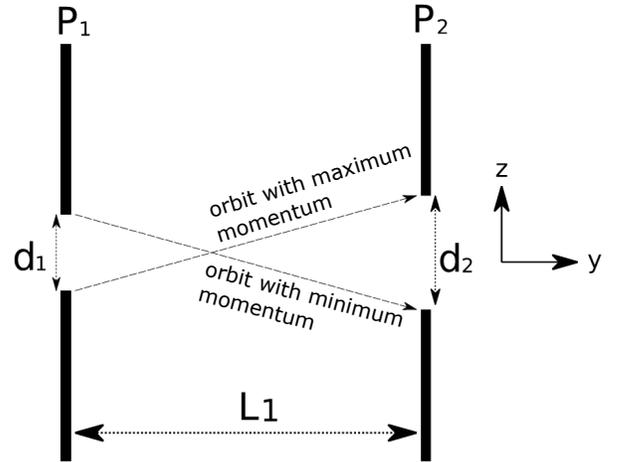}
\caption{Geometry of the collimator and the slit.}
\label{collimator}
\end{center}
\end{figure}

As depicted in Figure \ref{collimator} the parameters $D_p$ and $D_z$ are estimated by taking into account  
the half width $\de P$ of the possible classical momentum
after passing through the collimator (with plates 1 and 2) and the half width $\de Z$ 
of the possible classical position after passing through the slit (on plate 2) as 
\deq{
D_p&\sim \de P=\frac{d_1+d_2}{2L_1}mv_y,\\
D_z&\sim \de Z=\frac{d_2}{2}.
} 
To make unambiguous estimates, we suppose  that 
\deq{
0.75\,\de P\le D_p\le 1.25\,  \de P,\\
0.75\,\de Z\le D_z\le 1.25\, \de Z.
}

From \Eq{error} the error $\ep(\si_z)$ of the original Stern--Gerlach
measurement is given by
\deq{
\ep(\si_z)^2=2\,{\rm erfc}\left(\frac{g_0}{\sqrt{2}\si(\Det)}\right).
\label{eq:original-error}
}
Then, according to the parameter values given in Table \ref{SGexptable},
we have 
\deq{
0.972\le\dfrac{g_0}{\sqrt{2}\si(\Det)}\le 1.62,
\label{eq:EDR-ORG-1}
}
and, therefore, we conclude
\deq{
4.38\times 10^{-2}\le
\ep(\si_z)^2\le 3.38\times 10^{-1}.
\label{eq:error-org}
}
For the disturbance $\ep(\si_x)$ of the original Stern--Gerlach
measurement, from \Eq{disturbance} we have 
\deq{\et(\si_x)^2=2.
\label{eq:disturbance-org}}
See Appendix for the detailed calculations.

From the above we conclude that the error probability $\ep(\si_z)^2/4$ of the experiment 
is at most $8.5\%$.  This appears to be  consistent with Stern--Gerlach's original estimate of the
error to be  $10\%$ based on the agreement between the observed deflection and the
theoretical prediction \cite{GS22c}.

\begin{figure}[h]
\begin{center}
\includegraphics[width=0.45\textwidth]{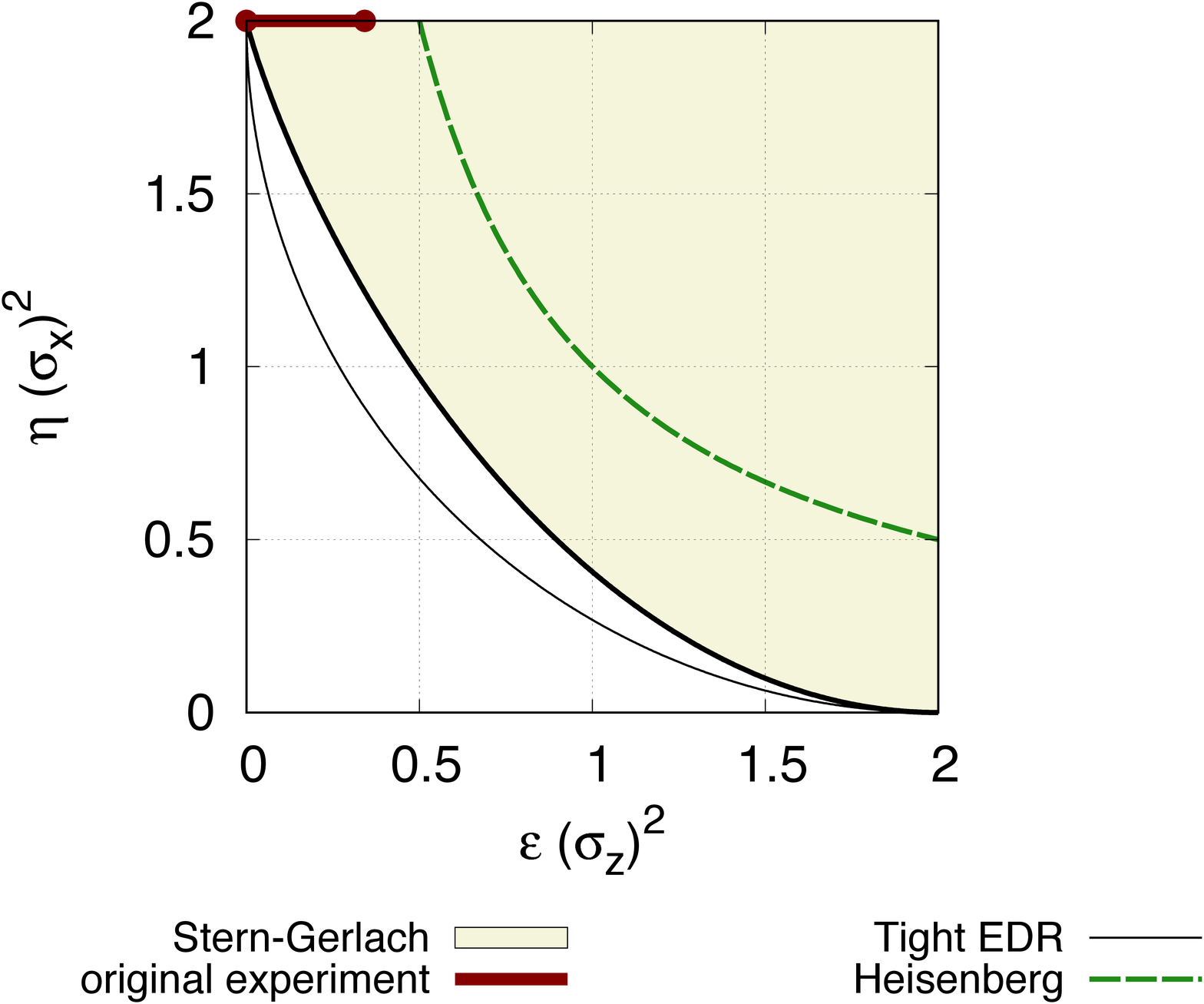}
\caption{
The estimated error--disturbance region 
for the original experiment performed by Gerlach and Stern \cite{GS22a,GS22b,GS22c} in 1922.
Beige region: the region \eq{EDR-SG} that Stern--Gerlach measurements can achieve.
Deep red line: the estimated error--disturbance region \eq{error-org}, \eq{disturbance-org}
for the original Stern--Gerlach  experiment in 1922. 
Black thine line: the boundary of 
the tight EDR \eq{BOEDRSZ}.
Green dashed  line: the boundary of Heisenberg's  EDR \eq{Hei27}.
}
\label{edcomp}
\end{center}
\end{figure}

As depicted in Figure \ref{edcomp}, the estimated error--disturbance region clearly violates  
Heisenberg's EDR.

\section{Conclusion}
In a previous study \cite{IO20}, we have determined the range of the error and disturbance 
taken by Stern--Gerlach measurements and compared it with the tight EDR for general spin 
measurements.  It is interesting to see that Stern--Gerlach measurements occupy
the near optimal subregion of the theoretically achievable region. 
Base on the above theoretical results, here, we have estimated the error 
and disturbance of the original Stern--Gerlach experiment performed in 1922,
and concluded that the original Stern--Gerlach experiment violates Heisenberg's
EDR.  This suggests that Heisenberg's EDR is more ubiquitously violated 
than we have believed for a long time, and it opens a new research interest 
exploring violations of Heisenberg's EDR in more common measurement setups 
to deepen our understanding of Heisenberg's uncertainty principle.
It will contribute to new developments in precision measurements 
such as optomechanical metrology and multi-messenger astronomy.

\acknowledgments
This work was partially supported by JSPS KAKENHI, Nos.~26247016 and 
17K19970, and the IRI-NU collaboration.

\bigskip

\section{
\bf Appendix: 
Derivations of Eq.~(\ref{eq:error-org}) and Eq.~(\ref{eq:disturbance-org})
}

From 2018 CODATA, the Boltzmann constant, the Avogadro constant $N_A$,
the electron magnetic moment $\mu$, and the reduced Planck constant $\hbar$ are given by
\deqs{
k_B&=1.380649\times 10^{-23} [{\rm J/K}],\\
N_A&=6.02214076\times10^{23} \left[{\rm mol^{-1}}\right],\\
\mu &=-9.2847647043\times  10^{-24} [{\rm J/T}],\\
\hbar&= 1.054571817\times 10^{-34}[{\rm J\cdot s}].
}
The mass $m$ of the silver atom with the standard atomic weight 
$107.86822[{\rm g/mol}]$ is given by
\deqs{
m &=\frac{1.0786822\times 10^{-1}[{\rm kg/mol}]}{6.02214076\times10^{23} \left[{\rm mol^{-1}}\right]}\\
&=1.7911939\times 10^{-25}[{\rm kg}].
}
From \Eq{v_y} and Table 1 we obtain 
\deqs{
v_y &=\sqrt{\frac{4k_B T}{m}}
=\sqrt{\frac{4\times 1.380\times 10^{-23}\times 1500}{ 1.791\times 10^{-25}}}\\
&= 6.80\times 10^2 [{\rm m/s}].
}
From Table 1 we obtain 
\deqs{
\Det& = \frac{L_2}{v_y} \\
&= 
\frac{3.5 \times 10^{-2}}{6.80\times 10^{2}} \\
&=5.14\times 10^{-5}[{\rm s}].
}

As depicted in Figure \ref{collimator},
the parameters $\de P$ and $\de Z$ are introduced as
\deqs{
\de P&=\frac{d_1+d_2}{2L_1}mv_y,\\
\de Z&=\frac{d_2}{2}.
} 
We obtain
\deqs{
1.25 \de Z &= \frac{5 d_2}{8} =2.50\times 10^{-6}[{\rm m}],\\ 
1.25 \de P &=\frac{5(d_1 + d_2)}{8L_1}mv_y\\
&=\frac{3.1\times 10^{-4}+2.0\times 10^{-4}}{8\times 3.3\times 10^{-2}}
\times 1.791\times 10^{-25}\\
&\quad\times 6.80\times 10^2\\
&=2.35\times10^{-25}[{\rm kg\cdot m/s}]. 
}
The parameters $D_p$ and $D_z$ are assumed to satisfy 
\deqs{
D_p&=1.25\, K{\de P},\\
D_z&=1.25\, K{\de Z}
}
for $0.6\le K\le 1$.
We obtain
\deqs{
{\rm Var}(Z,\xi_\la)
&=\frac{1}{4}\left(\frac{D_p^2}{\hbar^2}+\frac{1}{4D_z^2}\right)^{-1}\\
&=\frac{1}{4}\left(\frac{(K\times 2.35\times 10^{-25}[{\rm kg\cdot m/s}])^2}{(1.054\times 10^{-34}[{\rm J\cdot s}])^2}\right.\\
&\qquad\qquad+\left.\frac{1}{4(K\times 2.50\times10^{-6}[{\rm m}])^2}\right)^{-1}\nn\\
&=
\frac{1}{4}\left(K^2\times 4.97\times 10^{18}[{\rm m^{-2}}]\right.\\
&\qquad\qquad+\left.K^{-2}\times 4.00\times10^{10}[{\rm m^{-2}}]\right)^{-1}\\
&=K^{-2}\times 5.03\times 10^{-20}[{\rm m^{2}}],\\ 
\frac{\Det^2}{m^2}{\rm Var}(P,\xi_\la)&=\frac{\Det^2}{m^2}\frac{\hbar^2}{4{\rm Var}(Z,\xi_\la)}\\
&=\frac{(5.14\times 10^{-5})^2}{(1.791\times 10^{-25})^2}\\
&\qquad\qquad\times
\frac{(1.054\times 10^{-34})^2}{4\times K^{-2}\times 5.03\times 10^{-20}}\nn\\
&=K^2\times 4.54\times10^{-9} [{\rm m^2}],\\ 
\si(\Det)^2&={\rm Var}(Z,\xi_\la)+\frac{\Det^2}{m^2}{\rm Var}(P,\xi_\la)\\
&=\frac{\Det^2}{m^2}{\rm Var}(P,\xi_\la)\\
&=K^2\times  4.54\times10^{-9} [{\rm m^2}],\\ 
g_0&=\frac{\mu B_1 \Det^2}{2m}\\
&=\frac{(-9.28\times10^{-24}[{\rm J/T}])}{2\times (1.791\times 10^{-25} [{\rm kg}])}\\
&\qquad\times(-1.35\times 10^{3}[{\rm T/m}])\\
&\qquad\times(5.14 \times10^{-5}[{\rm s}])^2\\
&=9.26\times10^{-5}[{\rm m}],\\
\frac{g_0}{\sqrt{2}\si(\Det)}
&=\frac{9.26\times 10^{-5}}{K\sqrt{2\times4.54\times10^{-9}}}\\
&=K^{-1}\times 0.972.
}

From \Eq{original-error} we have
\deqs{
\ep(\si_z)^2&=2\,{\rm erfc}\left(\frac{g_0}{\sqrt{2}\si(\Det)}\right)\\
&=2\,{\rm erfc}\left(K^{-1}\times  0.972\right).}
For $K=1$, we obtain
\deqs{
2\,{\rm erfc}(0.972)&=2\times 0.1692=3.38\times 10^{-1}.}
For $K=0.6$, we obtain
\deqs{2\,{\rm erfc}(0.972/0.6)&=2\,{\rm erfc}(1.620)=2\times 0.0219\\
&=4.38\times 10^{-2}.
}
Thus, we conclude
\beqa
&0.972\le\dfrac{g_0}{\sqrt{2}\si(\Det)}\le 1.620,&\nn\\
&4.38\times 10^{-2}\le
\ep(\si_z)^2\le 3.38\times 10^{-1}.&\nn
\eeqa

To calculate the disturbance $\et(\si_x)$, we have
\deqs{
 \si\left(\frac{\Delta t}{2} \right)^2 
 &=\frac{1}{4} \si\left( \Delta t\right)^2\\
& =K^2\times 1.135\times10^{-9} [{\rm m^2}],\\ 
 \frac{\mu B_1 \Delta t} {\hbar}
&=  
\frac{
(-9.28\times10^{-24}[{\rm J/T}])}{1.054\times 10^{-34}[{\rm J\cdot s}]}\\
&\qquad\times(-1.35\times 10^{3}[{\rm T/m}])\\
&\qquad\times(5.14\times10^{-5}[{\rm s}] )\\
&=6.10\times 10^{9}[{\rm m}^{-1}],\\
 \frac{2 \mu ^2 B_1^2 \Delta t^2} {\hbar ^2}
\si\left( \frac{\Delta t}{2} \right)^2 
 &=2\times (6.10\times 10^{9}[{\rm m}^{-1}])^2\\
 &\qquad\times K^2\times 1.135\times10^{-9} [{\rm m^2}]\\
& =K^2\times 8.44\times 10^{10},\\ 
\lefteqn{
2\exp \!\left[
 -\frac{2 \mu ^2 B_1^2 \Delta t^2} {\hbar ^2}
 \si\left( \frac{\Delta t}{2} \right)^2
\right]}\quad\qquad\qquad\qquad\qquad\\
&=2\exp(-K^2\times 8.44\times 10^{10})\\
&=0.
 }
Thus, from \Eq{disturbance} we conclude
\begin{align*}
\eta(\sigma _x )^2
&=
2-2
\exp \!
\left[
 -\frac{2 \mu ^2 B_1^2 \Delta t^2} {\hbar ^2}
\si\left( \frac{\Delta t}{2} \right)^2
\right]
\cos \!
\frac{2\mu \Delta t B_0}{\hbar}\\
&=2. 
\end{align*}

\end{document}